# An Automated Laboratory Laser Heating Arrangement for Materials Synthesis at High Temperatures and High Pressures

N. Subramanian[*], Mrithula Ummru, R. Bindu, N. R. Sanjay Kumar, M. Sekar, N. V. Chandra Shekar, and P. Ch. Sahu

Advanced Materials Section, Materials Science Division,
Indira Gandhi Centre for Atomic Research, Kalpakkam-603102, Tamilnadu, India

[*] Corresponding Author:  Dr. N. Subramanian
Materials Science Division
Indira Gandhi Centre for Atomic Research,
Kalpakkam 603102, Tamil Nadu,
India

E-mail: subbi@igcar.gov.in

Phone: +91-44-27480347
Fax:    +91-44-27480081

**Short Title: Automated LHDAC for HP-HT Experiments**

**Abstract**

This paper describes the automation of a laser heating arrangement for synthesizing and studying materials at high pressures (up to ~ 1 Mbar) and high temperatures (up to ~ 5000 K). In this arrangement, a diamond anvil high-pressure cell (DAC) containing a microscopic sample of typical diameter ~50-100 μm, is mounted on a precision X-Y nanomotor stage that forms part of an IR laser heating optical assembly. Automation of this stage has been accomplished using a LabVIEW virtual instrument program to manipulate the X and Y stages using nanopositioning systems. This has a major feature of enabling a rastered heating of the sample over a user-defined circular area, without any operator intervention in addition to a virtual joystick to position the sample with respect to the laser spot. This auto-rastering feature has the advantage of offering uniform exposure of a circular area of the sample to the incident heating laser beam apart from drastic reduction in scan time compared to a manual scan. The diameter of the circle can be varied from a maximum of ~24 mm down to the focal spot size of the laser (~few μm), enabling thereby usage of the laser heating arrangement for heating microscopic samples under high-pressure in a DAC, as well as bulk samples at atmospheric pressure. Examples for both macro and micro scale automated laser-heating experiments have been presented. In particular, at the micro scale, auto-raster heated carbon samples at ~ 17 GPa and ~2000 K showed excellent signatures of diamond formation compared to manually raster heated samples, highlighting the unique advantage of auto-raster heating.

# 1. Introduction

Laser heating of microscopic samples contained in a Diamond Anvil Cell (DAC) has emerged in the recent years as an important route to synthesis and study materials at extremely high pressures and temperatures [1-10]. In typical experiments, using the Laser Heated Diamond Anvil Cell (LHDAC) technique, the material under study is placed in a small hole (diameter ~ 250 μm) drilled in a stainless steel gasket placed between two diamond anvils. An IR laser beam, either from $CO_2$ laser, Nd:YAG laser, or Nd:YLF laser, is focused on to the sample which has typical dimensions of ~ 70 to 150 μm to heat it. If, by accident or neglect, the laser beam is focused in any other region apart from the sample, there is a risk of damage to the diamond anvil cell. In order to avoid this as well as to achieve optimum focus of the laser beam on the sample, precision translation stages are usually employed. Single sided as well as double sided laser heating are in wide use presently in synchrotron beam lines and laboratory [11, 12]. The latter is superior in view of the reduction in the axial (i.e., DAC axis) temperature gradients in the sample. Both methods, however suffer from non-uniform heating along the radial direction. Usage of a multimode beam has helped in heating the sample over a wider area as compared to that possible with a Gaussian beam on account of almost uniform power distribution over larger spatial extent [12]. While this has served to reduce the radial temperature gradient, it suffers from the drawback that laser powers far higher than that of the Gaussian mode are needed to produce the same temperature effects.

In this article, we present details of automation of a single-sided laser heating arrangement that has been recently set-up in our laboratory [5]. The algorithm we have implemented enables the user to raster the sample with the heating laser beam over a desired circular area, thus bringing about uniform and systematic exposure of the entire sample to the laser spot. In our LHDAC facility [5], the DAC is mounted on a precision X-Y-Z-θ stage and is manipulated with respect to the static focused Gaussian $CO_2$ laser beam while viewing the hot spot on the sample surface with the help of a CCD imaging system. The Z-stage is a manually operated micrometer stage to vary the sample position along the incident laser beam direction and bring about effective focusing. θ-stage is used to rotate the sample about a vertical axis with respect to the horizontal laser beam, and is also manually operated. The X and Y stages are motorized and have minimum step resolution of 10 nm. A computer program, "AMS NANOMOVER" has been developed by us using LabVIEW (National Instruments, USA) Virtual Instrumentation Programming [13] to perform remote manipulation of the X and Y nanomover stages. RS-232 serial interfacing has been used for communication between the computer and the nanomotor controller. In heating experiments using this program, the operator has to manipulate the "virtual" joystick on the computer screen to raster the sample step by step to achieve reasonable uniform heating. This process requires attention of the operator at all time. Our experience has been that in some synthesis experiments, the heating may need to be done for couple of hours even, and can subject the operator to fatigue. Implementing a feature in the nanomover control routine, which enables automatic rastering of the sample, can help avert this, as also address the more important issue of uniform and complete heating across the sample face that can enhance the fraction of homogeneously reacted species.

# 2. Double-Pass Raster Heating Algorithm

Raster heating of a sample with the laser beam can be done by either moving the laser beam with respect to the static sample or by moving the sample with respect to the fixed laser beam. The former method requires coordinated X-Y movements of the focusing optics and requires maintaining of the focus over the entire scan area, which will be a non-trivial task. The latter method is relatively simpler because the optics is fixed and only the X-Y stage on which the sample is mounted needs to be manipulated. Moreover X-Y stages, in most set-ups like ours, have to anyway be coupled to precision motors to position the sample with respect to the incident beam and achieve optimum focusing. In this section, the algorithm that has been used by us to automate this latter method is discussed. Specifically the algorithm deals with moving the X-Y translation stages, coupled to nanomovers, in a synchronized manner with respect to the fixed and focussed $CO_2$ laser beam to achieve uniform heating of the sample over a circular area. The load bearing capacity of the nanomotor system used in our set-up is 10 Kg for horizontal movement (X) and 20 Kg for vertical movement (Y), while the DAC and stages weigh only about 5-6 Kg. This is well within limits and hence can not be expected to pose major problems during motion.

The circular rastering of the sample with respect to the laser beam can be achieved in several ways. One way would be to perform the scan along concentric circles, starting from, say, the center, and moving by Δr along the radius. The difficulty with this method lies in achieving a very high degree of synchronization between the X and Y stages which have to be operated alternating a number of times for each step. With heavy loads such as the stages and DAC, this will be impossible to realize in practice. An alternate way that overcomes this is to have a Y-step in the vertical direction executed first, followed by X-motion along the chord to one extreme of the circle. At that extremity, the next Y-step is executed followed by another X-motion in the reversed direction. This process goes on until the full circle is scanned. In both the scans described, heating takes place only once at any given point on the circular area, i.e., it amounts to single-pass heating per scan routine.

A third possibility that incorporates double-pass heating per scan routine is illustrated in Fig. 1a. In this mode, firstly Y-scan in the vertical direction starts from the center of the circle ($X_c$, $Y_c$) and ends at the top. From there, a small Y-step, ΔY is executed in the downward direction, along the diameter to reach a new position $Y_i$. Next, with respect to the new Y-position, an X-scan is done along the chord in either directions, and in the process enabling double-pass heating. After the execution of the requisite number of Y-steps, a full double-pass heating over the entire circular area would have been achieved. This scan routine, made clear in the algorithm given below, is also easily adaptable to the X-Y linear motion stages used in the laser-heating set-up.

If "D" be the diameter of the circular area to be scanned, "R" the radius, "ΔY" the step size and "$x_i$" the half-chord length (Fig. 1a), the number of steps (or number of chords), "N", to be scanned is,

N = D/ΔY　　　　　　　　　　　　　　　　　　　　　　　　　　---------- (1)

$x_i^2 + (R^2 - y_i^2) = R^2$,

$x_i = \sqrt{2Ry_i - y_i^2}$

$y_i = \sqrt{Dy_i - y_i^2}$

At the i$^{th}$ step, the distance "$y_i$" moved in the Y-direction is given by
$y_i = i\,\Delta Y$
Then,
$x_i = \sqrt{Di\Delta Y - i^2\Delta Y^2}$ , where "i" runs from 0 to N        --------- (2)

This algorithm has been implemented using LabVIEW [12]. Inputs to the circular scan program are the circle diameter, step size, and velocity of scan. The upper and lower limits of the velocity are 2500 μm/s and 5 μm/s respectively. A safe maximum input diameter of 24 mm can be input to the program as the nanomotors have a maximum range of 25 mm only.

## 3. Automated Laser Heating
### (a) Heating at Macroscopic Scale: Testing of the program and synthesis of large volume sample at high temperature and atmospheric pressure

Prior to testing the automation program with microscopic samples inside the high–pressure cell, it was imperative to test the algorithm extensively at a macroscopic scale of ~ few mm. The plan was to perform a laser-heating experiment involving synthesis of a known inorganic compound by heating uniformly a mixture of two parent compounds over a large circular area. The objectives were three-fold, namely, checking for compliance of the laser-engraved pattern with that expected from the algorithm, synchronization between the X and Y motions and finally the product formation.

The inorganic compounds chosen to be synthesized were BaFBr and BaFI, two technologically important matlockite structured materials that have been studied extensively by us earlier [14, 15, 16]. Primarily, the experiments comprised of attempting to laser-heat mixtures of the parent alkaline earth halides, $BaF_2$ and $BaX_2$ (X=Br, I), with the $CO_2$ laser and checking for compliance of the laser-engraved pattern with that expected from the algorithm, synchronization between the X and Y motions and product formation (BaFBr and BaFI) as a result of uniform heating.

The compounds $BaF_2$ and $BaX_2.nH_2O$ were mixed with a pestle and mortar and kept under an IR lamp for one hour to reduce the moisture content. Then the mixtures were compacted in a cavity of diameter 8 mm and depth 2 mm on an aluminum plate. It was then mounted on the X-Y-Z-θ stage (nanomotion system) and the $CO_2$ laser beam was focussed to fall on its center by manipulating the "virtual" joystick in the AMS-NANOMOVER program. The radiation from the hot sample was made to fall on a 100 μm diameter optical fiber and from there to a spectrometer for recording the thermal spectrum for temperature determination by spectroradiometry technique [5]. The heating was also viewed and recorded through an imaging system.

Prior to commencing the circular scan, the $CO_2$ laser spot was centered on the sample by manipulating the X-Y stages using the "virtual joysticks" that we have incorporated in the LabVIEW program. Next, for the circular-area raster scan, the diameter and step size were given as inputs. The mixed halides were circular scanned for a diameter of 5 mm, with step sizes of 500 μm and 250 μm for the $BaF_2+BaBr_2$ and $BaF_2+BaI_2$ mixtures respectively.

The $CO_2$ laser power was increased from 0 to 5% of its maximum power of 125 W, while viewing the sample on the monitor. This power was found to be just sufficient to induce melting. Using the circular raster scan program, heating was carried out on both the samples one after the other. The experimental conditions and the parameters input to the heating routine are given in Table 1.

During the initial run, it was found that the synchronization between the X and Y motions was getting lost. To start with, the heating proceeded in a synchronous manner and the sample-heating did proceed as per the circular-scan routine. However, after passage of some time, it was observed that the circle was getting increasingly distorted, on account of loss of synchronization due to the pile-up of commands over the serial interface as well as inertia effects. Loss of synchronization due to command pile-up arose due to the fact that the motion of a mechanical system occurs over a time scale which is much longer than that taken by the computer to execute the loops in the program. For instance when the X-motion is taking place in the $i^{th}$ chord, the computer almost instantaneously sends commands for moving the Y-motor to the $(i+1)^{th}$ step, X-motion along the $(i+1)^{th}$ step and so on. In fact, by the time a single X-motion along a chord is completed, all the commands for all subsequent scans would also have been dumped to the system by the computer, leading to total loss of synchronization.

An additional factor that can lead to loss of synchronization is mechanical in origin. As the nanomover moves with a weight of few kilograms of load (stages, sample holder, etc) with a high speed, it cannot suddenly change the direction of motion at the end of the chord as required by the program, owing to inertia. This, in turn will lead to loss of coordination between the X-motion and Y-motion. The minimum speed possible with the nanomotors, namely 5 μm/s, is still high enough to cause loss of synchronization.

Both the above factors were eliminated by introducing time-delays in the programming sequence at appropriate places to delay execution of commands at inappropriate moments. The command pile-up problem was eliminated, for instance, by making the Y-stage wait for a time "$t_i$", which was the same as the time taken to complete the double-pass scan along the "$x_i$" chord. Thus,

$$t_i = \frac{4x_i}{V}$$, where, V is the velocity of X-stage. The factor '4' occurs because each full chord scan comprises of a left motion by "$x_i$", a motion along the full chord $2X_i$ towards the right, and another motion to the left by "$x_i$", to bring the beam back to the starting position on the chord. A further delay of "$t_e$" seconds were added at the extremes of the chord-lengths and before commencement of next Y-step. Thus, the total delay that was introduced per step was,

$$T_{d,i} = 4t_e + \frac{4x_i}{V}$$

After implementing this delay timer in the LabVIEW program, it was seen that perfect synchronization in motion was achieved.

Figure 1b shows the photograph of the $BaF_2+BaBr_2$ sample taken after the heating scan. The pattern of channels cut by the laser beam in the circular area along the chord and the diameter can be seen to exactly resemble fig. 1a. During the raster-heating, the laser spot was observed to cut channels by melting the sample along the chords of the circle, precisely as desired from the scanning program. It can be seen that wherever melting has taken place, the sample appears glazed due to resolidification. The channels have an average width of ~30 μm and depth of ~ 1.5 mm. Measurement of the circular heated area under an optical microscope showed that the diameter was very close to 5 mm and there were 10 chords as expected. In Fig. 1b, a C-shaped channel can be seen in the top region. This is a result of another heating-scan programmed to inscribe a smaller circle ~ 2 mm diameter that was aborted at an early stage as the laser power was set too high. Laser heating of a mixture of $BaF_2+BaI_2$ too exhibited exactly the expected laser-engraved pattern.

In both the heating experiments, significant melting had taken place, as was seen by the spread of the molten region laterally in the spacing between the channels. One could certainly expect that fusing of the starting species in the mixture would have occurred at these temperatures leading to synthesis of the matlockites. Powder was carefully scrapped from various heated regions of the sample, ground in a pestle and mortar and XRD patterns were recorded using a Guinier diffractometer described in detail by Sahu et al [17].

Figure 2 shows the powder XRD pattern of one of the specimens scrapped from the heated $BaF_2+BaBr_2$ sample. Comparison of the pattern with the standard powder diffraction data file [18] of the matlockite BaFBr, represented as vertical lines immediately below, shows that they match very well. Analysis of the pattern using NBS[*]AIDS83 program [19] yielded lattice parameters a = 4.511(1) Å and c = 7.452(3) Å. These compare well with the lattice parameters published in the standard PDF file 70-1144 [18]. The figure of merit, $F_{15}$, of the fit was 34.9, which is extremely good. It may be pointed here that the figure of merit $F_N$, where N is the number of peaks observed and input to the analysis program, is given by [20] $F_N = \frac{N_T}{N}\left(\frac{1}{\Delta 2\theta_{Av}}\right)$, where, $N_T$ is the number of theoretically possible Bragg lines within the given θ-range, and $\Delta 2\theta_{AV}$ is the average FWHM of the Bragg peaks.

Likewise, XRD experiments confirmed formation of BaFI subsequent to laser heating the $BaF_2+BaI_2$ mixture. The time taken for carrying out the laser-heating on the 5000 μm diameter circular area is about 30-60 minutes, depending on the step size and velocity input to the program. Compared to the conventional solid-state reaction methods, where, the heating time is typically of ~few hours, this novel method is considerably faster and can be used to synthesize single-phase materials successfully. By choosing a step size slightly less than the focal spot diameter, the entire circular area can be rastered without channel formation to achieve perfectly uniform heating.

**(b) Heating at Microscopic Scale: Testing and synthesis experiments at high pressure and high temperature using a DAC**

An important test of the automated raster-scan will be to check it an actual high-pressure high-temperature experiment, in which the input parameters "D" and "ΔY" are scaled down to microscopic dimension. In these experiments, as mentioned earlier, a microscopic sample "squeezed" between the tips of two diamonds in a Diamond Anvil Cell (DAC) has to be raster-heated over a small circular area of typical diameter ~ 100 μm, with a focused laser spot of diameter ranging from ~5 μm (for Nd:YAG laser) to ~ 30 μm (for $CO_2$ laser). The weight of the DAC taken with the X-Y stages is ~5-6 Kg. Maintaining perfect synchronization between the X and Y motions for micron level motions in order to inscribe the circular area can therefore be expected to be a very difficult task due to inertia effects. Nevertheless, we set out to use the circular raster-heating program on a Ru+C mixture contained in a Mao-Bell type DAC in an attempt to synthesis ruthenium carbide, which is expected to be a high-strength solid that we have been working on recently [5, 21].

The sample chamber, which was a preindented stainless steel gasket having a 250 μm (diameter) x 80 μm (depth) hole, was first completely filled with a small quantity of NaCl and compressed in the DAC so that it became transparent. A shallow pit was then scooped out in the NaCl, in which the Ru+C sample in the form of a densified chip ~ 100 μm diameter was placed. Above this assembly a thin layer of NaCl was placed to ensure thermal insulation of the diamond. A ruby crystal ~10-15μm diameter was placed in the sample chamber for pressure calibration. The role of NaCl, which is an IR window, was to act as pressure transmitter apart from thermally insulating the diamonds from the hot spot.

Figure 3 depicts snapshots extracted from a video movie of a laser-heating experiment done at high-pressure on the Ru+C sample using the circular raster-heating program [22]. Scan circle diameter of 100 μm, step size of 10 μm, and scan speed of 50 μm were given as input to the program. The scan speed was deliberately kept high, in fact the same as for macroscopic scale experiment, in order to test the efficacy of the X-Y stages to execute the scan algorithm with heavy load of the DAC. The sample is seen as a circular disc at the centre in the background of transmitted light through the transparent NaCl present everywhere in the circular hole. The dark region outside this transparent region is the stainless steel gasket. The bright spot seen on the sample is the light emitted by it subsequent to absorption of the 10.6 μm $CO_2$ laser wavelength. Sample pressure, determined by ruby fluorescence technique [23] is ~ 8 GPa and the temperature ~ 2000 K. It can be seen that the average spot size is ~ 25 μm-30 μm and is very bright in the central region (frame 1) which is the home position for the circular scan. Once the scan commences (frames 2-8), the spot is seen to execute the pattern as per the double-pass heating algorithm. Further, the spot size is seen to decrease during motion due to the decreased dwell time at any given position.

Although, XRD of the heated sample did not show any evidence of any new phase formation, this experiment with the double-pass circular raster-heating routine shows that even for ultra small dimensions (Scan circle diameter ~ 100 μm, step size ~ 10 μm), heavy loads and high scan speeds, the program is working well and there is perfect synchronization between the X and Y stages till end-of-scan. The provision of adequate time delay between the various stages in the scanning algorithm seems to be effective in maintaining the synchronization of the X and Y motions even with the heavy DAC positioned on the stage.

Next, two experiments involving laser-heating of graphite present inside the DAC at ~17 GPa and ~ 2000 K were done. The first one involved manual heating at several locations on the 100 μm diameter sample, while the second one employed the novel automated raster-heating. Comparison of Raman spectra of both samples show excellent signature of $sp^3$-bonded phase (diamond) in the case of auto-raster-heated sample (Figure 4) highlighting the unique advantage of the auto-rastering over the manual heating. Appearance of the strong characteristic first order diamond peak at 1330 cm$^{-1}$ [24] in the raster-heated sample is a result of uniform exposure and heating of the sample inside the DAC to the IR laser. In fact, SEM studies done in conjunction with Raman studies also show a higher fraction of cubic granules in the raster-heated samples compared to the manually heated samples [25].

## 4. Conclusions

With the successful incorporation and testing of the circular raster-heating, the automated laser heating arrangement at our laboratory presents itself as a versatile tool for synthesizing bulk samples over several mm$^2$ of area at atmospheric pressure or for microscopic samples at varying high pressures and high temperatures. This has been exemplified by experiments involving successful synthesis of microscopic quantities of diamond from graphite at high pressure and temperature and synthesis of large volumes of BaFBr compound at atmospheric pressure and high temperature. A few possible features, which are currently under development by us are: (1) Ultra slow scan, i.e., scan rates that are less than the lowest velocity of 5 μm/s inherent to the X-Y nanomotors. This may be needed in cases where sample needs to be exposed for long time to laser spot before absorption commences or in cases when quenching rates need to be low. One way to achieve this is to consider the chord as a discrete array of slices and set the dwell time per slice to as long as needed. This way, even though during transit the nanomotors will move with 5 μm/s or higher speeds, the effective scan rate can be made much lesser. An added advantage of increasing the dwell time is that the heat will pervade deeper into the sample; (2) Implementation of a gradient in scan speed, as opposed to the present constant speed. This may be needed in experiments at very high pressures, when, due to large pressure gradient the sample may show a gradient in absorption of the laser beam as well. This can lead to non-uniform heating and can be overcome to some extent by either varying the laser power proportionately or by slowing down the sample scan rate in regions where absorption is less. The latter option is feasible to be implemented in the automation program that has been presented here.

The raster-heating program can be easily used in double side heating arrangements as well without requiring any modification. In laboratory as well as synchrotron beam lines, such an automated arrangement will prove advantageous to subject the sample to heating in a uniform manner in the radial directions.

**Acknowledgements**

Authors thank Shri. L. Meenakshi Sundaram for technical help. They thank Shri P. Chandramohan and Dr. M. P. Srinivasan for Raman spectroscopy of laser heated graphite. The constant encouragement and support rendered by IGCAR management is gratefully acknowledged.

**Table 1** Laser power and input parameters for circular area raster heating program.

| Sample | Laser power (Watts) | Scan circle diameter D (μm) | Step size ΔY (μm) | Scan Speed (μm/s) | Number of chords (D/ΔY) |
|---|---|---|---|---|---|
| $BaF_2$ + $BaBr_2$ | 6.3 | 5000 | 500 | 50 | 10 |
| $BaF_2$ + $BaI_2$ | 6 | 5000 | 250 | 50 | 20 |

**Figure Captions**

**Fig. 1.** (a) Schematic of double-pass laser-heating scan over circular area. The point ($X_c$, $Y_c$) defines the center of the circle of radius R; (b) Photograph of $BaF_2$+$BaBr_2$ sample after laser-heating using the algorithm developed presented in this work. The channels that have been cut by the laser are precisely as expected from the algorithm. The C-shaped cut seen in the top half of the heated-region is the result of an aborted heating run for a smaller circle radius.

**Fig. 2.** Powder XRD pattern of mixture of $BaF_2$+$BaBr_2$ after laser-heating. The standard pattern of BaFBr taken from the PDF file 70-1144 is also shown below for comparison. The peak marked "*" belongs to $BaF_2$.

**Fig. 3.** Raster laser-heating of a Ru+C mixture in a diamond anvil cell at ~ 8 GPa. Snapshots have been extracted from a video movie recorded of the entire circular scan.

**Fig. 4.** Raman spectra of laser heated graphite at ~ 14 GPa and ~ 2000 K. Spectrum (a) is that of the sample heated by manually rastering the laser spot; spectrum (b) corresponds to the sample heated by the automated rastering method. The peak at ~1330 $cm^{-1}$ is the characteristic first order (zone-centre optical phonon) Raman line of diamond [24].

**Fig. 1**

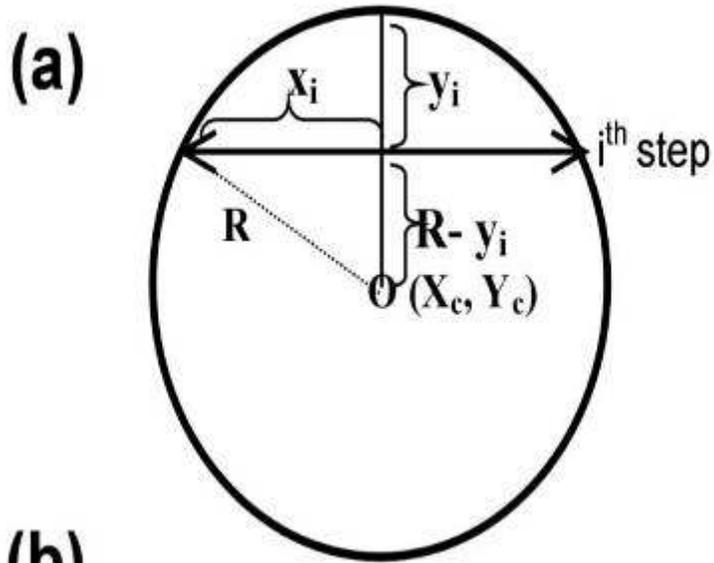

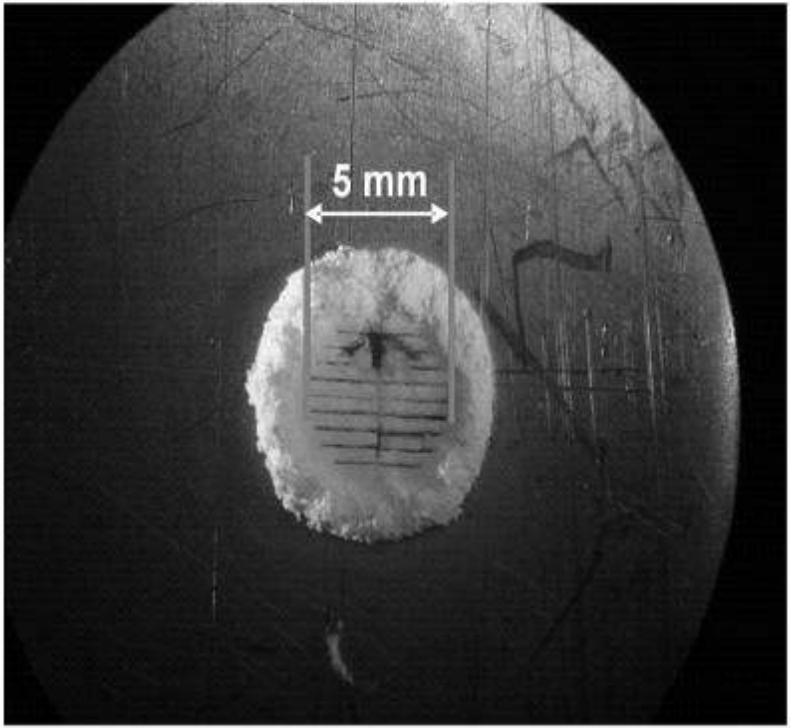

**Fig. 2**

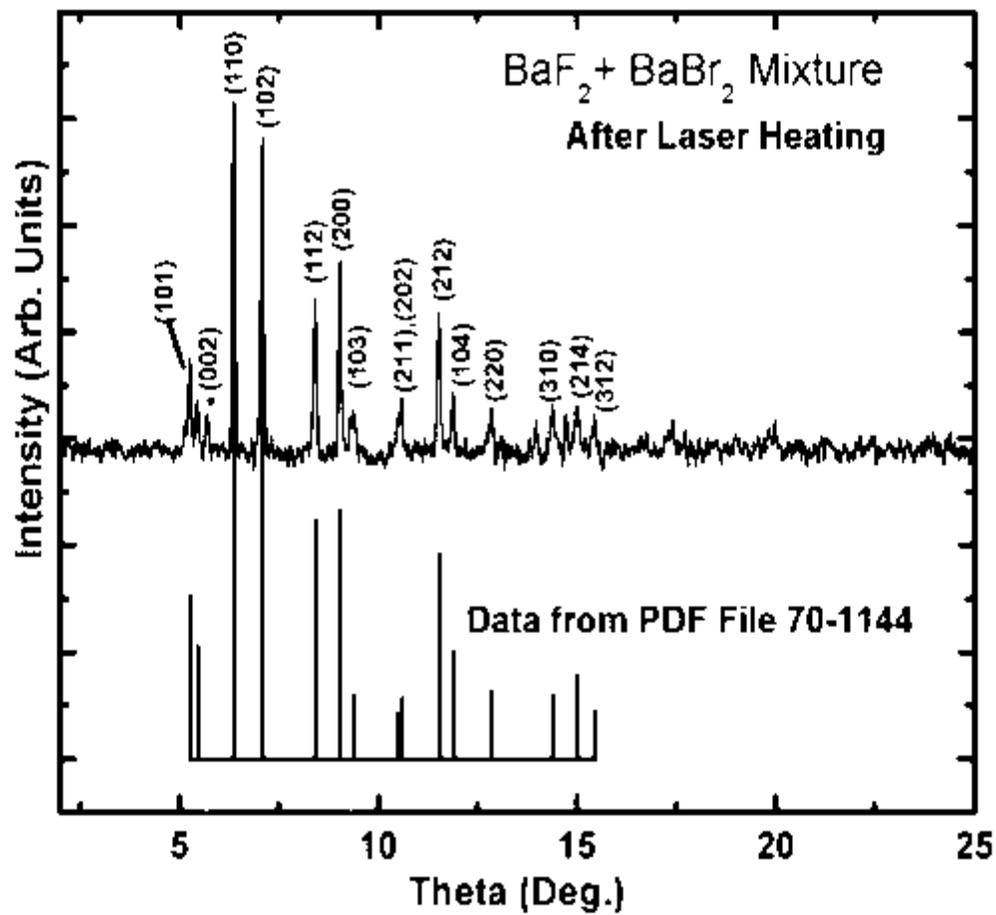

**Fig. 3**

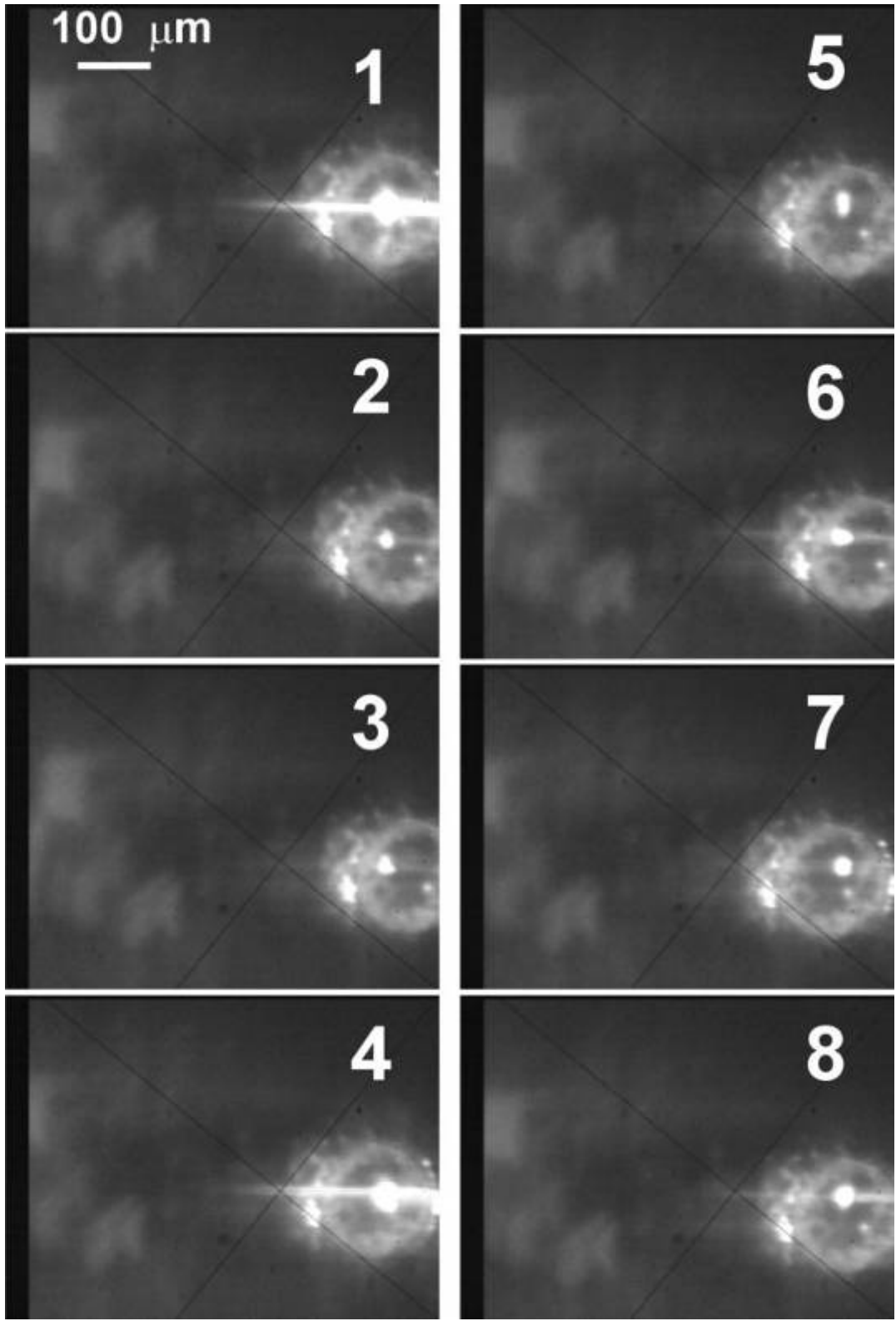

**Fig. 4**

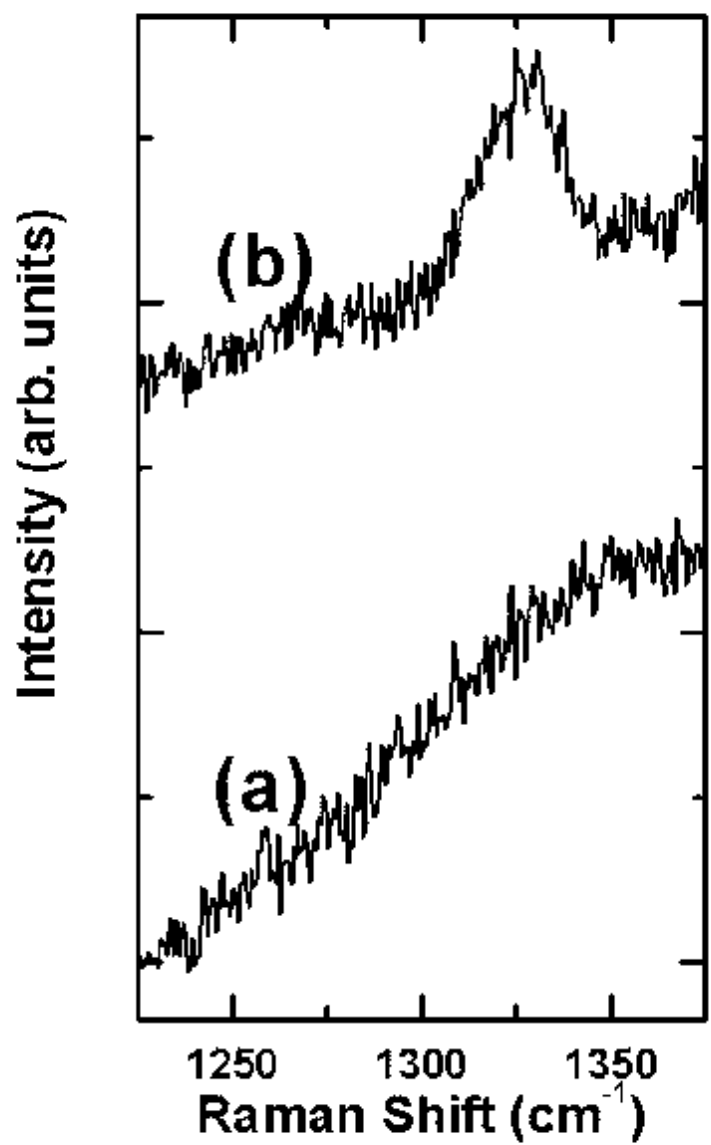